\documentclass[preprint,preprintnumbers,amsmath,amssymb]{revtex4}
\usepackage{amsfonts}    
\usepackage{amssymb}
\usepackage{latexsym}
\usepackage{graphicx}
\usepackage{rotating}
\usepackage{multirow}
\usepackage[usenames,dvipsnames]{color}
\usepackage[active]{srcltx}
\usepackage{setspace}
\usepackage{hyperref}
\usepackage{color}

\newcommand{\De}{\Delta}

\newcommand{\half}{\frac{1}{2}}
\newcommand{\rar}{\rightarrow}

\begin{document}

\title{Power-like potentials: from the Bohr-Sommerfeld energies to exact ones}

\date{\today}

\author{J.C.~del~Valle}
\email{delvalle@correo.nucleares.unam.mx}
\author{Alexander~V.~Turbiner}
\email{turbiner@nucleares.unam.mx}
\affiliation{Instituto de Ciencias Nucleares, Universidad Nacional
Aut\'onoma de M\'exico, Apartado Postal 70-543, 04510 M\'exico,
CDMX, Mexico{}}
\begin{abstract}
For one-dimensional power-like potentials $|x|^m, m > 0$ the Bohr-Sommerfeld Energies (BSE) extracted explicitly from the Bohr-Sommerfeld quantization condition are compared with the exact energies. It is shown that for the ground state as well as for all positive parity states the BSE are always above the exact ones contrary to the negative parity states where BSE remain above the exact ones for $m>2$ but they are below them for $m < 2$. The ground state BSE as the function of $m$ are of the same order of magnitude as
the exact energies for linear $(m=1)$, quartic $(m=4)$ and sextic $(m=6)$ oscillators but relative deviation grows with $m$ reaching the value 4 at $m=\infty$. For physically important cases $m=1,4,6$ for the $100$th excited state BSE coincide with exact ones in 5-6 figures.

It is demonstrated that modifying the right-hand-side of the Bohr-Sommerfeld quantization condition by introducing the so-called {\it WKB correction} $\gamma$ (coming from the sum of higher order WKB terms taken at the exact energies) to the so-called exact WKB condition one can reproduce the exact energies. It is shown that the WKB correction is small, bounded function $|\gamma| < 1/2$ for all $m \geq 1$, it is slow growing with increase in $m$ for fixed quantum number, while it decays with quantum number growth at fixed $m$. For the first time for quartic and sextic oscillators the WKB correction and energy spectra (and eigenfunctions) are written explicitly in closed analytic form with high relative accuracy $10^{-9-11}$ (and $10^{-6}$).
\begin{center}
	\begingroup
	\setstretch{1}
	\textit{Published version of this manuscript}: \\
	J.C. del Valle and A.V. Turbiner, \textit{Int. J. Mod. Phys. A} \textbf{36}, 29, 2150221 (2021). \\
	\href{https://doi.org/10.1142/S0217751X21502213}{https://doi.org/10.1142/S0217751X21502213}
	\endgroup
\end{center}
\end{abstract}

\maketitle

In semiclassical consideration of the quantum mechanics the celebrated Bohr-Sommerfeld quantization rule,
\begin{equation}
\label{BS}
  \frac{1}{\hbar} \int_{B}^{A} \sqrt{E - V(x)} d x\ =\ \bigg(N + \half\bigg) \pi\ +\ O(\hbar) \ ,
\end{equation}
see e.g. \cite{LL}, where $A,B$ are the turning points at the energy $E=V(A,B)$ of the $N$th excited state for the potential $V(x)$, plays very important role. It is a common knowledge that at larger $N$ the Bohr-Sommerfeld energy $E_{BS}$, extracted from (\ref{BS}), gets more and more accurate. This property is fundamental for making estimates on the highly-excited states. Surprising fact, which one of the authors (A.T.) assigns to Carl Bender \cite{Bender:1990} \footnote{See e.g. \cite{Bender-Orszag:1978} for the potential $x^4$, Table 10.4}, is that even the ground state energy, found by using (\ref{BS}), can be sufficiently accurate. The goal of this Note is two-fold: (I) to study the Bohr-Sommerfeld energy spectra $E_{BS}$ for one-dimensional power-like potentials
\begin{equation}
\label{V}
  V \ =\ a\, g^{m-2}\,|x|^m \ ,\ m>0\ ,
\end{equation}
where $a$ is dimensionless parameter (without loss of generality $a=1$), $g$ is the coupling constant coming from the anharmonic oscillators, where $x^2$-term is added to (\ref{V}), without a loss of generality $g=1$ \footnote{It can be shown that the Schr\"odinger equation with potential (\ref{V}) is one-parametric, it depends on the combination $g^2\,\hbar$, see e.g.  \cite{ST:2021}. Thus, the energy is $E=E(g^2\,\hbar)$.}, and then to compare $E_{BS}$ with accurate energies obtained in the Lagrange Mesh method \cite{Baye:2015,Tur-delValle:2021} for quantum numbers $N \in [0, 100]$, in this case the turning points $A,B=\pm E^{1/m}$ in (\ref{BS}); and (II) to modify the Bohr-Sommerfeld quantization condition (by introducing the so-called WKB correction, see below) up to the so-called {\it exact} WKB condition and obtain the modified Bohr-Sommerfeld energy spectra which should coincide with exact energy spectra
\footnote{For even $m > 2$ the potentials (\ref{V}) describe anharmonic oscillators at ultra-strong coupling regime, where the dependence on the effective coupling constant $g^2\,\hbar$.}.
These potentials naturally occur in celebrated ODE/IM correspondence \cite{Dorey-Tateo:1999}, for the general review see \cite{Dorey:2020}. Needless to say that the power-like potential (\ref{V}) is non-exactly-solvable in general (except for $m=2$ and $\infty$, where the spectra is found explicitly by algebraic means, and $m=1$, where finding the eigenvalues is reduced to solving a transcendental equation) and the eigenstates can be found approximately {\it only}. From the quantum field theory perspective the quantum mechanics with power-like potentials resembles massless scalar field theory.

The potential $V$ (\ref{V}) is characterized by quite exceptional property: the Bohr-Sommerfeld quantization condition (\ref{BS}) can be solved analytically with respect to $E$ leading to the BSE in the explicit form
\begin{equation}
\label{EBS}
		E_{BS}\ =\ \bigg({\hbar}\,M\,B\left(\tfrac{1}{2},M\right)\,
		\left(N+\frac{1}{2}\right)\bigg)^{\frac{1}{M}}\ ,
		\ M=\frac{1}{m}+\frac{1}{2}\ ,\ N=0,1,2,\ldots \ ,
\end{equation}
where $B(a,b)$ is the Euler Beta function, see e.g. \cite{Voros:1999} and references therein, and, in particular, \cite{Dorey-Tateo:1999}, Eqs.(3)-(4) therein \footnote{Note that there exists another potential $V = \log |x|$ for which the BSE can be found explicitly
$E_{BS}\ =\ \log \left[\sqrt \pi (N+\frac{1}{2})\right]$, see \cite{Gesztesy:1978}.
Interestingly, for all positive parity states $E_{BS} > E_{exact}$, while for negative parity states $E_{BS} < E_{exact}$.}. Contrary to that claimed before, the BSE for positive and negative parity state are given by the {\it same} function. Interestingly, although BSE are defined for discrete values of quantum number $N=0,1,2,\ldots$, they can be analytically continued to $N \in [0,\infty)$. It is naturally to assume that is true for the exact energies.

It can be immediately checked that at large $N$ the functional behavior of $E_{BS}$ energies coincides with the well-known asymptotic expansion of the exact energy spectra,
\begin{equation}
\label{1/N}
      E_{exact}\ = \ c_M\ N^{\frac{1}{M}}\,(1 +\ \frac{a}{N}\ +\ \ldots ) \ ,
\end{equation}
where
\[
      c_M\ =\ c^{(BS)}_M\ =\ \bigg({\hbar}\,M\,B\left(\tfrac{1}{2},M\right)\,
		\bigg)^{\frac{1}{M}}\ ,\ a^{(BS)}\ =\ \frac{m}{m+2}\ .
\]
In general, $a=a(m)$ is weakly-dependent on $m$ function: for small $m$ the function $a$ is close $a^{(BS)}$, for $m=2$ $a=a^{(BS)}=1/2$, while at large $m\rar\infty$ the function $a^{(BS)}$ tends to 1 and when $a$ for the exact energy spectra tends to 2. 

It is well known that
for the harmonic oscillator $m=2$ the BSE coincide with exact ones,
\begin{equation}
\label{EBSm=2}
		E_{BS}(m=2)\ =\ {\hbar}\,\left(2N+1\right)\ =\ E_{exact}\ ,
\end{equation}
In the limit $m \rar \infty$ the potential (\ref{V}) becomes symmetric square-well potential of width 2 with infinite walls and
\begin{equation}
\label{EBSm=infty}
   E_{BS}\     =\ \frac{\hbar^2 \pi^2\,(N+\tfrac{1}{2})^2}{4}\quad ,\quad
   E_{exact}\ =\ \frac{\hbar^2 \pi^2\,(N+1)^2}{4} \ .
\end{equation}
It is worth noting that replacing $1/2$ in the right-hard-side of (\ref{BS}) by $1$ the BSE coincide with exact ones.

In order to have a reference point we carried out detailed numerical calculations of energy spectra for three values of $m=1,4,6$ in (\ref{V}) in the Lagrange Mesh Method (LMM), employing the code designed for carrying out calculations in \cite{Tur-delValle:2021} and \cite{delValle}, with 100 mesh points \footnote{This code allows easily to reach accuracy in energies 200 figures (or more), see \cite{Tur-delValle:2021}}. With such a small number of mesh points it provided not less than four exact decimal digits in energy, which was sufficient for the most of this paper. These accuracies \footnote{Or higher ones, up to 10-11 significant digits} can be reached variationally, using the {\it Approximant} introduced in \cite{delValle} and taking the first two terms in Non-linearization Procedure \cite{Turbiner:1980-4}. Each particular calculation in LMM for given $m$ and $N$ took a few seconds of CPU time in a standard laptop.
Absolute deviation (A.D.) and relative deviation (R.D.) in energies are defined as
\begin{equation}
\label{A+R.D}
\text{A.D.}\ =\ E_{exact}-E_{\text{BS}}\quad ,\quad \text{R.D.}\ =\  \frac{\left|E_{exact}-E_{\text{BS}}\right|}{E_{exact}}	\ ,
\end{equation}
respectively. From now on we set $\hbar=1$.

\subsection*{Linear Symmetric Potential: $m=1$}

The results of calculations are presented in Table I. Exact energies $E_{exact}$ obtained in LMM are compared with $E_{BS}$ (\ref{EBS}) at $m=1$ and also with results from fit
\begin{equation}
	E_{fit}^{(|x|)}\ =\ 1.770\,683 \left(0.036\,284 + 0.984\,940 N + 1.228\,065 N^2
                        + 2.000\,000 N^3 + N^4\right)^{1/6} \ ,
\end{equation}
which agrees with $E_{exact}$ with relative accuracy $10^{-4}$ (or better). The first two leading terms in $1/N$ expansion for $E_{fit}$ and $E_{BS}$ coincide with six decimal digits
\begin{align*}
	E_{fit}^{(|x|)}\  &=\  1.770\,683 N^{2/3}\  +\    0.590\,227N^{-1/3}\  + \ \ldots \ ,\\
	E_{BS}^{(|x|)}\ &=\  1.770\,683N^{2/3}\ +\  0.590\,227N^{-1/3}\  + \ \ldots \ .
\end{align*}
Note that at $N=100$ the relative deviation $E_{exact}$ from $E_{BS}$ is of the order of $10^{-6}$. It signals that the difference $a^{(exact)}$ from $a^{(BS)}$ and $a^{(fit)}$ in (\ref{1/N}) is of the order $10^{-4}$.

The interesting observation, see Table I, is that for positive parity states $N=0,2,4,\ldots$ the exact energies are always smaller than the BSE contrary to the negative parity states $N=1,3,5,\ldots$ where the exact energies are always larger than the BSE. Similar behavior remains true for all potentials with $0 < m < 2$.

\begin{table}[h]
	\caption{Linear Potential: $V(x)=|x|$. For $E_{exact}$ all printed digits exact.}
	{\setlength{\tabcolsep}{0.4cm}		
\begin{tabular}{cccccc}
			\hline				
		  $N$ & $E_{exact}$ & $E_{fit}$ & $E_{BS}$ (3) & A.D. & R.D.
\\
			\hline		
			0 & 1.0188  & 1.0188  & 1.1155   & $-9.7\times 10^{-2}$  & $9.5\times10^{-2}$             \\
			1 & 2.3381  & 2.3344  & 2.3203   & $1.8\times 10^{-2}$   & $7.6\times10^{-3}$              \\
			2 & 3.2482  & 3.2625  & 3.2616   & $-1.3\times 10^{-2}$  & $4.1\times10^{-3}$             \\
			3 & 4.0879  & 4.0800  & 4.0818   & $6.1\times 10^{-3}$   & $1.5\times10^{-3}$            \\
			4 & 4.8201  & 4.8240  & 4.8263   & $-6.2\times 10^{-3}$  & $1.3\times10^{-3}$             \\
			5 & 5.5206  & 5.5148  & 5.5172   & $3.4\times 10^{-3}$   & $6.2\times10^{-3}$             \\
		   10 & 8.4885  & 8.4890  & 8.4905   & $-2.0\times 10^{-3}$  & $2.4\times10^{-4}$            \\
	       15 & 11.0085 & 11.0066 & 11.0077  & $8.0\times 10^{-4}$   & $7.3\times10^{-5}$           \\
		   20 & 13.2622 & 13.2623 & 13.2630  & $-8.3\times 10^{-4}$  & $6.2\times10^{-5}$              \\
		   25 & 15.3408 & 15.3397 & 15.3403  & $4.4\times10^{-4}$    & $2.9\times10^{-5}$               \\
		   50 & 24.1916 & 24.1916 & 24.1918  & $-2.5\times10^{-4}$   & $1.0\times10^{-5}$             \\
		   75 & 31.6306 & 31.6303 & 31.6305  & $1.0\times10^{-4}$    & $3.3\times10^{-6}$      \\
		  100 & 38.2752 & 38.2752 & 38.2753  & $-1.0\times10^{-4}$   & $2.6\times10^{-6}$             \\
			\hline				
\end{tabular}
	}
\end{table}

\subsection*{Quartic Oscillator: $m=4$}

The results of calculations are presented in Table II, cf. Table 10.4 in \cite{Bender-Orszag:1978}. Exact energies $E_{exact}$ obtained in LMM are compared with $E_{BS}$ (\ref{EBS}) at $m=4$ and also with results of the fit
\begin{equation}
	E_{fit}^{(x^4)}\  =\ 2.185\,069 \,\left(0.114\,266 + 0.529\,819 N + 1.623\,712 N^2 +
                         1.999\,492 N^3 + N^4\right)^{1/3}\ ,
\end{equation}
which agrees with $E_{exact}$ with relative accuracy $10^{-4}$ (or better). The first two leading terms in $1/N$ expansion for $E_{fit}$ and $E_{BS}$ coincide within six-three decimal digits before rounding, respectively,
\begin{align*}
    E_{fit}^{(x^4)}\  &=\   2.185\,069N^{4/3}\  +\   1.456\,343 N^{1/3}\  + \ \ldots \ ,\\
    E_{BS}^{(x^4)}\ &=\  2.185\,069N^{4/3}\ +\  1.456\,713N^{1/3}\  + \ \ldots \ .
\end{align*}
Like for the case of linear potential at $N=100$ the relative deviation $E_{exact}$ from $E_{BS}$ is of the order of $10^{-6}$, see Table II. It signals that the difference $a^{(exact)}$ from $a^{(BS)}$ and $a^{(fit)}$ in (\ref{1/N}) is of the order $10^{-4}$.

From Table II one can see that the exact energies are always {\it larger} than the BSE for both positive and negative parity states.

\begin{table}[h]
	\caption{Quartic potential: $V(x)=x^4$. For $E_{exact}$ all printed digits exact.}
	{\setlength{\tabcolsep}{0.4cm}		
		\begin{tabular}{cccccc}
			\hline				
			$N$ & $E_{exact}$ & $E_{fit}$& $E_{BS}$ (3) & A.D. & R.D. \\
			\hline		
			0   & 1.0604     & 1.0603  & 0.8671     & $1.9\times 10^{-1}$     & $1.8\times10^{-1}$             \\
			1   & 3.7997     & 3.8018  & 3.7519     & $4.8\times 10^{-2}$     & $1.3\times10^{-2}$              \\
			2   & 7.4557     & 7.4519  & 7.4140     & $4.2\times10^{-2}$      & $5.6\times10^{-3}$             \\
			3   & 11.6447    & 11.6434 &  11.6115   & $3.3\times10^{-2}$      &  $2.9\times10^{-3}$            \\
			4   & 16.2618    & 16.2615 & 16.2336    & $2.8\times10^{-2}$      & $1.7\times10^{-3}$             \\
			5   & 21.2384    & 21.2386 &  21.2137   & $2.5\times10^{-2}$      & $1.2\times10^{-3}$             \\
			10  & 50.2563    & 50.2570 &  50.2402   & $1.6\times10^{-2} $     &  $3.2\times10^{-4}$            \\
			20  & 122.6046   & 122.6050 & 122.5943  & $1.0\times10^{-2}$      & $8.4\times10^{-5}$              \\
			50  & 407.8744   & 407.8738 & 407.8687  & $5.7\times10^{-3}$      & $1.4\times10^{-5}$             \\
	       100 & 1020.9900   & 1020.9887 & 1020.9864 & $3.6\times10^{-3}$     & $3.5\times10^{-6}$\\
			\hline				
		\end{tabular}
	}
\end{table}

\subsection*{Sextic Oscillator: $m=6$}

The results of calculations for $m=6$ in (\ref{V}) are presented in Table III. Exact energies $E_{exact}$ obtained in LMM are compared with $E_{BS}$ (\ref{EBS}) and also with results of the fit
\newpage
\[
   E_{fit}^{(x^6)}\ =\ 2.265\,089\,\times
\]
\begin{equation}
	\left(0.065\,250 + 1.403\,063 N + 0.480\,511 N^2 +
                      3.433\,389 N^3 + 4.077\,198N^4+3.000\,334 N^5 + N^6\right)^{1/4}	\ ,
\end{equation}
which agrees with $E_{exact}$ with relative accuracy $10^{-4}$ (or better) similarly to $m=1,4$ cases. The first two leading terms in $1/N$ expansion for $E_{fit}$ and $E_{BS}$ coincide within six-three decimal digits, respectively,
\begin{align*}
	E_{fit}^{(x^6)}\  &=\ 2.265\,089N^{3/2}\ +\ 1.699\,006N^{1/2}\   + \ \ldots \quad ,\\
	E_{BS}^{(x^6)}\ &=\ 2.265\,089N^{3/2}\ +\  1.698\,817N^{1/2}\ + \ \ldots \quad .
\end{align*}
Similar to the case of linear potential $m=1$ and quartic potential $m=4$ at $N=100$ the relative deviation $E_{exact}$ from $E_{BS}$ is of the order of $10^{-6}$, see Table III. It signals that the difference $a^{(exact)}$ from $a^{(BS)}$ and $a^{(fit)}$ in (\ref{1/N}) is of the order $10^{-4}$.

From Table III one can see that the exact energies are always {\it larger} than the BSE like 
it was for quartic potential, see Table II. This property continues to hold for larger integer $m > 6$. In general, for any real $m > 2$ the exact energies are {\it larger} than the BSE. We guess that this property remains when the oscillator (\ref{V}) is extended to the anharmonic oscillator.

\begin{table}[h]
	\caption{Sextic Potential: $V(x)=x^6$. For $E_{exact}$ all printed digits exact.}
{\setlength{\tabcolsep}{0.4cm}		
	\begin{tabular}{cccccc}
\hline				
		$N$ & $E_{exact}$ & $E_{fit}$& $E_{BS}$ (3) & A.D. & R.D. \\
\hline		
		0   & 1.1448     & 1.1448    & 0.8008     & $3.4\times 10^{-1}$ & $3.0\times 10^{-1}$             \\
		1   & 4.3386     & 4.3385    & 4.1612     & $1.8\times 10^{-1}$ & $4.1\times 10^{-2}$              \\
		2   & 9.0731     & 9.0737    & 8.9536     & $1.2\times 10^{-1}$ & $1.3\times 10^{-2}$             \\
		3   & 14.9352    & 14.9341   & 14.8316    & $1.0\times 10^{-1}$ & $6.9\times 10^{-3}$            \\
		4   & 21.7142    & 21.7139   & 21.6224    & $9.2\times 10^{-2}$ & $4.2\times 10^{-3}$             \\
		5   & 29.2996    & 29.2998   & 29.2166    & $8.3\times 10^{-2}$ & $2.8\times 10^{-3}$             \\
		10  & 77.1273    & 77.1275   & 77.0671    & $6.0\times 10^{-2}$ & $7.8\times 10^{-4}$            \\
		20  & 210.2835   & 210.2833  & 210.2404   & $4.3\times 10^{-2}$ & $2.0\times 10^{-4}$              \\
		50  & 812.8999   & 812.9001  & 812.8725   & $2.7\times 10^{-2}$ & $3.4\times 10^{-5}$             \\
		100 & 2282.1182  & 2282.1189 & 2282.0988  & $1.9\times 10^{-2}$ & $8.5\times 10^{-6}$
\\
\hline				
	\end{tabular}
}
\end{table}

\subsection*{R.D. as a function of $m$}

\begin{figure}[h]
	\centering
	\includegraphics[]{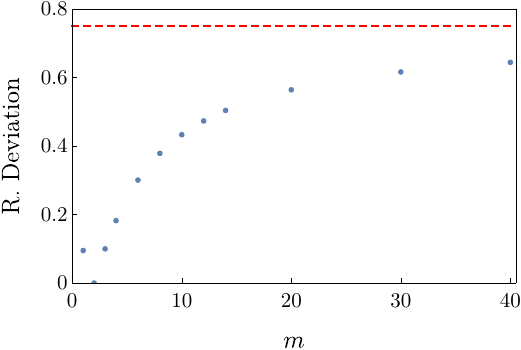}
\caption{R.D. of the Bohr-Sommerfeld ground state energy (\ref{BS}), $N=0$,
         from the exact one for potential $V(x)=|x|^{m}$ {\it vs} $m$. Points correspond to $m=1,2,3,4,6,8,10,12,15,20,30,40$\,. Maximal deviation 3/4 (dashed curve) reached when $m\rar\infty$.}
\label{N=0}
\end{figure}

Accurate energies $E_{exact}$ obtained in LMM, see, in particular, Tables I-III, can be compared with the BSE   $E_{BS}$ (\ref{EBS}) in order to estimate the relative deviation (\ref{A+R.D}) versus $m, N$. In Fig.\ref{N=0} the R.D. (\ref{A+R.D}) is presented for the ground state $N=0$ as the function of $m$ in (\ref{V}). For $m=2$ these energies coincide, thus R.D.=0, and then R.D. begins to grow for both $m<2$, when $m$ is decreasing, and $m>2$, when $m$ is increasing, reaching 3/4 at $m \rar \infty$, see (\ref{EBSm=infty}). Similar behavior occurs for $N=1,2,\ldots$ with monotonous decrease of R.D. for fixed $m$. Interestingly, for $N=100$ the R.D. is of the order $10^{-5}$ for small $m=1,4,6$ but reaching $\sim 10^{-2}$ at $m \rar \infty$.

\subsection*{How to improve BS energies to exact energies}

The right-hand-side of the Bohr-Sommerfeld quantization condition (\ref{BS}) is written in leading order in the Planck constant $\hbar$. In general, it contains infinite asymptotic expansion in powers of $\hbar$ which is not clear how to sum it up. However, it seems clear that, as the result of summation for fixed energy spectra, we should arrive at some function $\gamma$ of $m,N,\hbar$ and, implicitly, the energy,
\begin{equation}
 \frac{1}{\hbar} \int_{-E_{exact}^{1/m}}^{E_{exact}^{1/m}}\sqrt{E_{exact}-|x|^m}\,dx\ =\ \pi\left(N+\frac{1}{2}+\gamma\right)\ ,
\label{correctedBS}
\end{equation}
cf.(\ref{BS}), where $\gamma$ is introduced to the r.h.s., see e.g. \cite{Voros:1999} and references therein; we will call $\gamma$ the {\it WKB correction}. In general, $\gamma$ is related to the boundary condition imposed at turning point, it was claimed in \cite{LL} that $(1/2+\gamma)$ is of order one. In fact, Eq.(\ref{correctedBS}) can be considered as a definition of $\gamma$ when the energies in the l.h.s. are exact. As a result the explicit formula (\ref{EBS}) for the Bohr-Sommerfeld spectra is modified in a very simple, straightforward  manner
\begin{equation}
\label{EBS-modified}
		E^{(modified)}_{BS}\ =\ \bigg({\hbar}\,M\,B\left(\tfrac{1}{2},M\right)\,
		\left(N+\frac{1}{2}+\gamma(m,N;\hbar)\right)\bigg)^{\frac{1}{M}}\ .
\end{equation}
It must be emphasized if $\gamma$ in (\ref{correctedBS}) depends on the energy {\it explicitly} the formula (\ref{EBS-modified}) is not valid. Hence, our actual goal is to check the energy independence of the WKB correction $\gamma$, thus, the validity of (\ref{EBS-modified}). In two particular cases, when $m=2$ the function $$\gamma(2,N;\hbar)\ =\ 0\ ,$$ for any $N$ and when $m=\infty$ the function $$\gamma(\infty,N;\hbar)\ =\ 1/2\ ,$$ see (\ref{EBSm=infty}), the function $\gamma$ is known exactly and does not depend on energy. Needless to say that if $E_{exact}=E_{BS}$ the WKB correction is absent, $\gamma=0$.
Hence, the problem of finding the energy spectra of a power-like potential (\ref{V}) is reduced to
finding the WKB correction $\gamma$. Sometimes, the condition (\ref{correctedBS}) is called the
{\it exact} WKB quantization condition.

Making comparison of the accurate $E_{exact}$ at fixed $m,N$ with $E^{(modified)}_{BS}$ one can find unambiguously the function $\gamma(m,N;\hbar)$ but within the accuracy with which $E_{exact}$ is known.
Since LMM allows to find $E_{exact}$ with practically any accuracy we like, we can find $\gamma(m,N;\hbar)$ with a similar accuracy \footnote{Maximal accuracy in energies we had explored was 50 significant digits, typically, it required $\sim 2000$ mesh points.}.
On Fig.\ref{N=0,1} it is shown the behavior $\gamma_0=\gamma(m,N=0;\hbar)$ for the ground state,\ $N=0$, and $\gamma_1=\gamma(m,N=1;\hbar)$ for the first excited state, $N=1$,
versus $m$ in (\ref{V}). In both cases $\gamma$ demonstrates smooth, slow-changing, monotonously-growing behavior with increase of $m$ leading eventually to 1/2. For larger $N$ the behavior remains similar. Moreover, it can be shown that for $m > 2$ fixed if $N_1 < N_2$ it is always $\gamma(N_1) > \gamma(N_2)$. It is evident that the function $\gamma$ is non-negative and bounded, $\gamma(m,N;\hbar) \leq 1/2$, at least for $m \geq 2$.

\begin{figure}[h]
	\centering
	\includegraphics[]{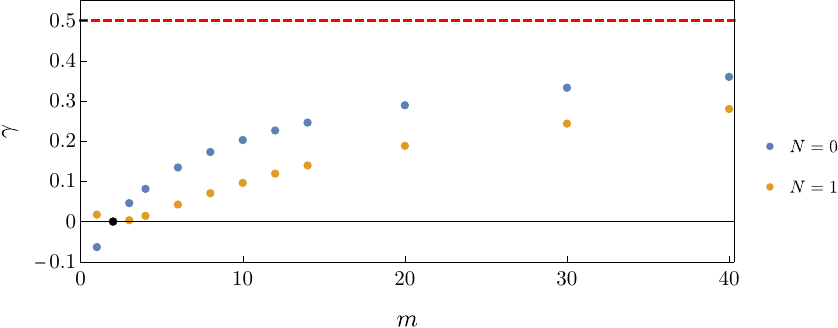}
	\caption{\label{N=0,1} WKB corrections $\gamma_{0,1}$ {\it vs} $m$,
              see (\ref{correctedBS}), for the first two states, $N=0,1$. Displayed
              points correspond to $m=1,2,3,4,6,8,10,12,15,20,30,40$\,. At $m \rar \infty$ both sequences of points tend asymptotically to 1/2 (shown by dashed curve).}
\end{figure}

Note that the function $\gamma(m,0;\hbar)$ can be easily fitted
\begin{equation}
      \gamma(m,0;\hbar)\ =\ \frac{\mu}{2\,m}\,\frac{0.0031\,\mu^2\ +\ 0.1041\,\mu\  +\ 0.2087 }{0.0031\,\mu^2\  +\ 0.1557\,\mu\ +\ 1}\ ,\qquad \mu\ =m-2\ .
\label{fit}
\end{equation}
This fit, when substituted to (\ref{EBS-modified}), allows us to reproduce at least 3 s.d. in ground state energy in domain $1 \leq m \leq 40$ for power-like potentials (\ref{V}). On Fig.\ref{m=4,6} the functions $\gamma(m=4,N)$ and $\gamma(m=6,N)$  for quartic and sextic potentials, respectively, are shown.
On Fig.\ref{m=1} the functions $\gamma(m=1,N)$ for even (upper sequence of dots) and odd (lower sequence of dots) states are shown. For all three cases $m=1,4,6$ the WKB correction $\gamma$ at large $N$ decays like $1/N$.

\begin{figure}[h]
	\centering
	\includegraphics[]{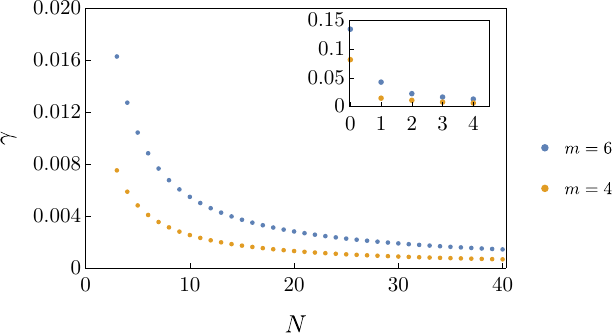}
	\caption{\label{m=4,6} WKB correction $\gamma$ {\it vs} $N$ for quartic $m=4$ and
     sextic $m=6$ potentials, see (\ref{EBS-modified}). Displayed points correspond to $N=0,1,2,3,\ldots, 40$\,. In subFigure the domain $N=0,1,2,3,4$ shown separately. }
\end{figure}
We are not aware how to calculate $\gamma$ versus $N$ at fixed $m$ analytically. Thus, it can be done numerically by solving the equation (\ref{EBS-modified}) backward with respect to $\gamma$ and assuming that $E^{(modified)}_{BS} = E_{exact}$, and then looking for relevant interpolating function in $m$. As the first approach this program was realized for quartic and sextic oscillators:
both sequences of points are shown in Fig.3.
They can be easily fitted via the ratio of a polynomial to square-root of another polynomial, 
\[
    \frac{P_n(N)}{\sqrt{Q_{2n+2}(N)}}
\]
where $P_n,Q_{2n+2}$ are polynomials \footnote{For reasons which are unclear to the present authors, contrary to the Pade approximants $P(n|n+1)$, this form does not lead to singularities at positive $N$.}. For example, the simplest approximation at $n=1$,
\begin{equation}
\label{m=4}
\gamma_{fit}^{(x^4)}(N) \ =\ \frac{0.0265\,  N\ +\ 0.0416}{\sqrt{N^4\ +\ 5.6459\,N^3\ -\ 6.2201 N^2\ +\ 22.0334\,N\ +\ 0.2616}}\ ,
\end{equation}
\begin{equation}
\label{m=6}
	\gamma_{fit}^{(x^6)}(N) \ =\ \frac{0.0574\,  N \ +\ 0.0520}{\sqrt{N^4\ +\ 3.4958\,  N^3\ +\ 2.078 \,  N^2\ -\ 0.0374\,  N\ +\ 0.1497}}\ .
\end{equation}
where $N=0,1,2,\ldots$ \ ,
lead to sufficiently high accuracy for quartic and sextic oscillators. 
These fits (\ref{m=4}),(\ref{m=6}), when substituted to (\ref{EBS-modified}), allow us to reproduce, at least, 4 decimal digits in energy for quantum numbers $0 \leq N \leq 100$. It implies that all numbers for $E_{exact}$ printed in Tables II,III in the first column \footnote{except for $N=3$ where for unclear reasons 3 decimal digits are reproduced only} are reproduced exactly.
As for the case $m=1$ the fit should be done separately for even and odd eigenstates,
\begin{equation}
\label{m=1+}
    \gamma_{fit}^{(|x|)}(N)_+\ =\ -\frac{0.0394\, N\ +\ 0.0505}{\sqrt{N^4\ +\ 3.5543\, N^3\ +\ 4.8344\, N^2\ +\ 2.9564\,N\ +\ 0.6321}}\ ,\quad N\ =\ 0,2,4,\ldots \ ,
\end{equation}
\begin{equation}
\label{m=1-}
	\gamma_{fit}^{(|x|)}(N)_-\ =\ \frac{0.0281\,N+0.0567}{\sqrt{N^4\ +\ 5.0343\, N^3+8.7658\,N^2\ +\ 7.1521\,N\
    +\ 1.9661}}\ ,\quad N\ =\ 1,3,5,\ldots \ ,
\end{equation}
respectively, cf.(\ref{m=4}),(\ref{m=6}). After substitution of (\ref{m=1+}), (\ref{m=1-}) to (\ref{EBS-modified}) all four decimal digits printed in Table I for energy, the first column, are reproduced. We are certain that similar fit performed for other values of $m$ will lead to results of the similar quality.
\begin{figure}[h]
	\centering
	\includegraphics[]{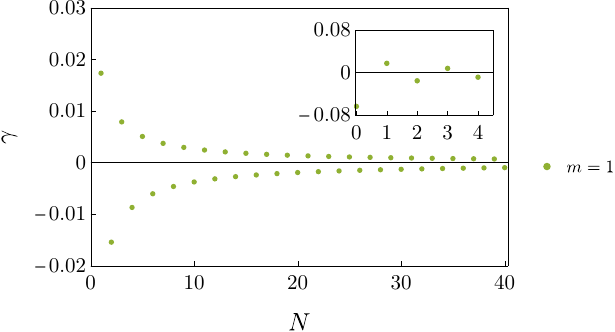}
	\caption{\label{m=1} WKB correction $\gamma$ {\it vs} $N$ for $m=1$ potential,
          see (\ref{EBS-modified}). Displayed points correspond to $N=0,1,2,3,\ldots, 40$\,.
          Even states correspond to upper sequence of points and odd states related with lower sequence of points. }
\end{figure}

The accuracy in $\gamma$ as well as of the fit of $\gamma$ can be increased if more accurate values of energies are taken as entries in (\ref{correctedBS}). In particular, for quartic oscillator the energies found with 8 exact significant digits allow us to find $\gamma$'s with 8 significant digits, see for consistency (\ref{EBS-modified}). They are fitted with
\begin{equation}
\label{m=4m}
   \gamma_{fit}^{(x^4)}(N) \ =\ \frac{0.026525\,N\ +\ 0.001203 }{\sqrt{N^4\ +\ 1.457179\, N^3\ -\ 1.792243\, N^2\ +\ 3.094978\, N\ +\ 0.000218}} \ ,
\end{equation}
cf.(\ref{m=4}), where $N=0,1,2,3,\ldots$\,. However, further increase in accuracy of description of energies will require to use a modified fitting function for $\gamma$,
\begin{equation}
\label{m=4m_2-6}
   \gamma_{fit}^{(modified)}\ =\ \frac{P_2(N)}{\sqrt{Q_6(N)}}\ ,
\end{equation}
where $P_2,Q_6$ are polynomials of degrees 2 and 6, respectively. Additionally, the increase in accuracy beyond of 8 significant digits reveals a new feature of the quartic oscillator: the WKB corrections for positive and negative parity states, calculated in (\ref{EBS-modified}), are (slightly) different and should be fitted separately. In particular, in order to get 10 significant digits in energy accurately the fit should be
\begin{equation}
\label{fit_m=4+_10}
	\gamma^{(x^4)}_{fit}(N)_+\ =\
\end{equation}
\begin{equation*}
\frac{0.02652582\,N^2 - 0.03913219\,N + 0.08142236}
{\sqrt{
N^6 - 1.95122872 N^5 + 5.84890038 N^4 - 2.38726524 N^3 + 5.55021093 N^2 + 3.98746583N
+ 1.22015178 
}}
\end{equation*}	

\noindent
for positive parity states, $N=0,2,4,\ldots$\,,

\begin{equation}
\label{fit_m=4-_10}
\gamma^{(x^4)}_{fit}(N)_-\ =\
\end{equation}
\begin{equation*}
 \frac{0.02652582\, N^2 + 0.01178769\, N - 0.00534128}
 {\sqrt{N^6 + 1.88728273 N^5 + 1.17139608 N^4 - 0.12919918N^3  - 0.17756539 N^2 + 1.29044237 N
 + 0.27714859}}
\end{equation*}

\noindent
for negative parity states, $N=1,3,5,\ldots$\,, cf.(\ref{m=4}),(\ref{m=4m}). Eventually, the formula (\ref{EBS-modified}) at $m=4$ with  $\gamma^{(x^4)}_{fit}(N)_{\pm}$ (\ref{fit_m=4+_10})-(\ref{fit_m=4-_10}) provides not less than exact 10 significant digits in energy.

Similar situation occurs for sextic oscillator: in order to get accuracy beyond of 8 significant digits the positive and negative parity states should be fitted separately. In particular, in order to get 9 significant digits in energy accurately the fit for the WKB correction $\gamma$ should be

\begin{equation}
\label{fit_m=6+_9}
\gamma^{(x^6)}_{fit}(N)_+\ =\
\end{equation}
\begin{equation*}
	\frac{0.13449783 - 0.13783412N +
		0.05743009N^2}{\sqrt{1.00000011 + 4.54042200N - 3.75241407N^2 -
			2.16026844N^3 + 5.98149752N^4 -
			3.79952593N^5 + N^6}} \ ,
\end{equation*}	
for $N=0,2,4,\ldots$\ ,

\begin{equation}
\label{fit_m=6-_9}
\gamma^{(x^6)}_{fit}(N)_-\ =\
\end{equation}
\begin{equation*}
	\frac{0.03360701 - 0.05559154N +
		0.05743009N^2}{\sqrt{0.00831615 - 2.97840199N +
			3.96675141N^2 - 1.06973081N^3 +
			0.71874670N^4 - 0.94436250N^5 + N^6}} \ ,
\end{equation*}	
for $N=1,3,5,\ldots$\,. It can be considered as the indication to validity the formula (\ref{EBS-modified}), in other words, to the absence of the explicit energy dependence on $\gamma$.

\section*{Conclusions}

Concluding we have to state that we have all reasons to believe that the modified Bohr-Sommerfeld formula (\ref{EBS-modified}) leads to the {\it exact} energy spectra
\[
		E_{exact}\ =\ \bigg({\hbar}\,M\,B\left(\tfrac{1}{2},M\right)\,
		\left(N+\frac{1}{2}+\gamma(m,N;\hbar)\right)\bigg)^{\frac{1}{M}}\ ,
		\ M=\frac{1}{m}+\frac{1}{2}\ ,
\]
for the potential
\[
       V(x)\ =\ |x|^m \ ,
\]
for any $m>0$ once the appropriately-accurate energy-independent function $\gamma=\gamma(m,N;\hbar)$ is supplied. This guess was checked for the first 101 eigenstates $N=0,1,2,\ldots, 100$ for two important particular cases:
 
\noindent
(I) for quartic oscillator $(m=4)$ where the energy spectra is given by
\begin{equation}
\label{m=4exact}
		E^{(m=4)}_{exact}\ =\
           \bigg(\frac{3}{4}\,{\hbar}\,B\left(\tfrac{1}{2},\tfrac{3}{4}\right)\,
		\left( N+\frac{1}{2}+\gamma^{(4)} \right)\bigg)^{\frac{4}{3}}\ ,
\end{equation}
with $\gamma^{(4)}=\gamma(4,N;\hbar)$ coming from the interpolations (\ref{m=4}),(\ref{m=4m}) and (\ref{fit_m=4+_10})-(\ref{fit_m=4-_10}), see Fig.3; they provided relative accuracies in energy of 4 decimal digits, 8 significant digits and eventually, 10 significant digits, respectively, only in the last case of 10 significant digits the difference in $\gamma$'s for positive and negative parity states occurs, for accuracies of 8 significant digits (or less) these $\gamma$'s coincided, and

\noindent
(II) for the sextic oscillator $(m=6)$, where the energy spectra is given by
\begin{equation}
\label{m=6exact}
		E^{(m=6)}_{exact}\ =\
           \bigg(\frac{2}{3}\,{\hbar}\,B\left(\tfrac{1}{2},\tfrac{2}{3}\right)\,
		\left( N+\frac{1}{2}+\gamma^{(6)} \right)\bigg)^{\frac{3}{2}}\ ,
\end{equation}
with $\gamma^{(6)}=\gamma(6,N;\hbar)$ as the interpolation presented in (\ref{m=6}) leading to 4 significant digits in the energy spectra and 9 significant digits if the interpolations (\ref{fit_m=6+_9})-(\ref{fit_m=6-_9}) are used, see for illustration Fig.3, similarly to quartic case $\gamma$'s coincide for the accuracy of 8 significant digits or less, for higher accuracies $\gamma$'s are different for positive and negative parity states.
  
Now, the important question arises about eigenfunctions which lead to $E^{(m=4,6)}_{exact}$ (\ref{m=4exact}), (\ref{m=6exact}). In \cite{AHO} the following functions for the $[N=(2n+p)]$-excited state with quantum numbers $(n,p)$, $n=0,1,2,\ldots\ ,\ p=0,1$\,, were presented for quartic oscillator $V=x^4$,
\[
 \Psi^{(n,p;m=4)}_{(approximation)}\ =\
 \frac{x^p P_{n,p}(x^2)}{\left(B^2\ +\ x^2 \right)^{\frac{1}{4}}
 \left({B}\ +\ \sqrt{B^2\ +\ x^2} \right)^{2n+p+\frac{1}{2}}}
\]
\begin{equation}
\label{functions-4}
   \exp \left(-\ \dfrac{A\ +\ B^2\,x^2/6\ +\ x^4/3}
  {\sqrt{B^2\ +\ x^2}} \ +\ \frac{A}{B}\right)\ ,
\end{equation}
where $A=A_{n,p},\ B=B_{n,p}$ are two free parameters, and
\[
\Psi^{(n,p;m=6)}_{(approximation)}\ =\
\frac{x^p\,P_{n,p}(x^2)}
{(D^2\ +\ C^2 x^2\ +\ x^4 )^{\frac{1}{4}}\,
(D\ +\ \sqrt{D^2\ +\ C^2\,x^2\ +\ x^4 })^{n+\frac{p}{2}+\frac{1}{4}}
}
\]
\begin{equation}
\label{functions-6}
\times\ {\exp \left(-\dfrac{A\ +B\,x^2\ +\ C^2x^4/8\ +\ x^6/4}{\sqrt{D^2\ +\ C^2\,x^2\ +\ x^4 }}\ +\ \dfrac{A}{D}\right)} \ ,
\end{equation}
for sextic oscillator $V=x^6$, where $A=A_{n,p},\ B=B_{n,p}, \ C=C_{n,p},\ D=D_{n,p},$ are four free parameters. Here $P_{n,p}$ in (\ref{functions-4}) and (\ref{functions-6}) are some polynomials of degree $n$ in $x^2$ with positive roots, which are found by imposing the orthogonality condition to the states with quantum numbers less than $n$.
It was shown that taking (\ref{functions-4}),(\ref{functions-6}) as trial functions with free parameters as variational, the variational energies are obtained with unprecedental relative accuracy $\sim 10^{-10}$ (or better) or $\sim 10^{-12}$ (or better), respectively, 
\footnote{In particular, for quartic oscillator if $A_{0,0}=-1.8028,\ B_{0,0}=2.1470$ the variational energy $E_{var}=1.060 362 092$, while the accurate energy obtained in LMM is $E_{LMM}=1.060 362 090$: their absolute deviation $A.D.=2. 10^{-9}$. For sextic oscillator if $A_{0,0}=-3.2816,\ B_{0,0}=0.5831, C_{0,0}=4.0195,\ D_{0,0}=3.0146$ the variational energy $E_{var}=1.144 802 453 803$, while the accurate energy obtained in LMM is $E_{LMM}=1.144 802 453 797$: their absolute deviation $A.D.=6. 10^{-12}$.}
and the relative deviation of the trial functions (\ref{functions-4}),(\ref{functions-6}) from exact ones is $\sim 10^{-6}$ in any point in $x$-space. It is evident that the expectation values of the appropriate Hamiltonians over $\Psi^{(n,p;m=4,6)}_{(approximation)}$ with variationally optimized parameters are equal to the modified Bohr-Sommerfeld energies $E^{(m=4,6)}_{exact}$ with appropriately accurate $\gamma$'s, with relative accuracy $\sim 10^{-10}$ . Hence, the WKB corrections $\gamma$'s can be written via the variational parameters in (\ref{functions-4}),(\ref{functions-6}).

Note that for the linear potential $m=1$ the spectra is given by
\begin{equation}
\label{m=1exact}
		E^{(m=1)}_{exact}\ =\
           \bigg(\frac{3}{2}\,{\hbar}\,B\left(\tfrac{1}{2},\tfrac{3}{2}\right)\,
		\left( N+\frac{1}{2}+\gamma^{(1)}_{\pm} \right)\bigg)^{\frac{2}{3}}\ ,
\end{equation}
with $\gamma^{(1)}_{\pm}=\gamma_{\pm}(1,N;\hbar)$, where signs $(\pm)$ correspond to positive/negative states, coming from the interpolations (\ref{m=1+}), (\ref{m=1-}), respectively, see Fig.4. These interpolations leads to the energy spectra with accuracy of 4 decimal digits. 


\vspace{-5mm}
\section*{Acknowledgments}
\vspace{-3mm}

The authors thank C.M.~Bender, M.A.~Shifman, R.~Tateo, M.~Znojil for their interest to the work and useful discussions and, especially, G.~Dunne for sharing the profound knowledge of the subject.
This work is partially supported by CONACyT grant A1-S-17364 and DGAPA grant IN113819 (Mexico).


\newpage
\appendix
\begin{center}
	\section*{Quantum WKB correction: Addendum to \\ \textit{Power-like potentials: from the Bohr-Sommerfeld energies to exact ones}}
\end{center}
\begin{center}
	($\Omega$Dated: June 20, 2025)
\end{center}
\begin{center}
	\textit{This Addendum refers to the publication:} \\ 
	J.C. del Valle and A.V. Turbiner, \textit{Int. J. Mod. Phys. A} \textbf{36}, 29, 2150221 (2021).
	\href{https://doi.org/10.1142/S0217751X21502213}{{https://doi.org/10.1142/S0217751X21502213}}
\end{center}

\subsection*{}
	\textbf{Abstract}. Numerical exploration suggests that for power-like potentials $|x|^m$ at $m>0$ the quantum WKB correction $\gamma$, introduced in \cite{BS-1:2021}, has the form $\gamma=\gamma(I^2)=|I|^{-1} {\tilde \gamma} (I^{-2})$, where $I=N+1/2$ is the {\it shifted} quantum number $N$. The leading coefficient in the $1/I$-expansion is found to be a simple analytic expression, while several subleading terms are found numerically with high accuracy. 
\vskip0.5cm

During a seminar at Stony Brook University, Prof. N. Nekrasov conjectured \cite{NN:2025} that the ``right" variable for calculating the quantum WKB correction $\gamma$ in the {\it exact} quantization condition
\begin{equation*}
	\frac{1}{\hbar} \int_{-E_{exact}^{1/m}}^{E_{exact}^{1/m}}\sqrt{E_{exact}-|x|^m}\,dx\ =\ \pi\left(N+\frac{1}{2}+\gamma\right)\ ,
	\label{correctedBS}
\end{equation*}
see \cite{BS-1:2021}, is the shifted in 1/2 quantum number,
\begin{equation}
	\label{Nshifted}
	I\ =\ N\ +\ \frac{1}{2}\ .
\end{equation} 
The goal of this short paper, which can be considered as Addendum to \cite{BS-1:2021}, 
is to verify this conjecture numerically in different circumstances.

{\bf 1.}\ Let us take an approximation of $\gamma$ of the form
\[
\gamma\ =\ \frac{P_2(N)}{\sqrt{Q_6(N) }}\ ,
\]
see Eq.(19) in [1], where $P_2, Q_6$ are polynomials in the quantum number $N$ of degrees 2 and 6, respectively. 
This approximation was successfully employed for the potentials $|x|^{1,4,6}$ leading to relative accuracies in $\gamma$ of the order $10^{-6}$ in domain $N \in [0, 200]$. Undoubtedly, we can rewrite those polynomials $P_2, Q_6$ in terms of the shifted quantum number $I$ (\ref{Nshifted}),
\begin{equation}
	\label{g26}
	\gamma\ =\ \frac{{\tilde P}_2({I})}{\sqrt{{\tilde Q}_6({I}) }}\ ,
\end{equation}
where, evidently, ${\tilde P}_2, {\tilde Q}_6$ remain polynomials of the same degrees as $P_2, Q_6$.
It can be immediately seen for both the $x^4$ (see Eqs.(20-21) in [1]) and the $x^6$ (see Eqs.(22-23) in [1]) potentials, that the coefficients in front of the terms of odd degrees in ${I}$ of the polynomials ${\tilde P}_2, {\tilde Q}_6$ are very small numbers. This suggests setting these coefficients equal to zero, thus, to assume that the polynomials ${\tilde P}_2, {\tilde Q}_6$ are formally even: 
${\hat P}_1({I}^2)\equiv {\tilde P}_2(I)\,,\, {\hat Q}_3({I}^2)\equiv {\tilde Q}_6(I)$, and then to refit $\gamma$, obtained by using the first 201 eigenvalues in all three potentials $|x|^{1,4,6}$, see Eq.(12) in [1], which we were able to find reliably in the Lagrange Mesh method \cite{Mathematica} by using about 2000 mesh points with $\geq 200$ significant digits. 
In this case the fitting function (\ref{g26}) becomes,
\begin{equation}
	\label{g13}
	\gamma\ =\ \frac{{\hat P}_1({I}^2)}{\sqrt{{\hat Q}_3({I}^2)}}\ .
\end{equation}
This fitting procedure was {\it unexpectedly} done separately for the positive/negative parity states, where the quantum number $N$ is even/odd, respectively. Eventually, this results in
\[
\gamma^{(|x|)}_{+} =\frac{-0.039\,403 I^2-0.046\,881}{\sqrt{I^6+2.691\,230 I^4+1.901\,750 I^2+0.137\,250}}\ ,\] 
\begin{equation}
	\label{gamma-lin-pot}
	\gamma^{(|x|)}_{-} = \frac{0.028\,145 I^2+0.055\,765}{\sqrt{I^6+4.433\,122 I^4+5.414\,409I^2+1.102\,802}}\ ,
\end{equation}  
for the linear potential $|x|$, and 
\begin{equation}
	\label{gamma-quartic-pot}
	\gamma^{(x^4)}_{+} = \frac{\dfrac{1}{12 \pi} I^2 + 0.024\,575}{I^2 \sqrt{I^2 + 2.100\,243}}\ , \qquad
	\gamma^{(x^4)}_{-} = \frac{\dfrac{1}{12 \pi} I^2 - 0.002\,675}{\sqrt{I^6 + 4.511\,343}}\ ,
\end{equation}
and
\[
\gamma^{(x^6)}_{+} =
\frac{\dfrac{5}{16 \sqrt{3} \pi} I^2 - 0.014016}
{\sqrt{I^6 - 0.478\,854 I^4 + 2.653\,365 I^2 - 0.649\,032}}\ , 
\]
\begin{equation}
	\label{gamma-sextic-pot}
	\gamma^{(x^6)}_{-} =
	\frac{\dfrac{5}{16 \sqrt{3} \pi} I^2 - 0.002\,064}
	{\sqrt{I^6 + 2.100\,205 I^2 - 7.091\,050}}\ , 
\end{equation}
for positive/negative parity states for the $|x|^{4,6}$ potentials, respectively, 
{here the subscripts $\pm$ correspond to the positive/negative parities respectively}.
As a result, even with a smaller number of fitted parameters, we are able to improve 
the quality of the fit by increasing the number of correct decimal digits in $\gamma$ from 6 to 10-11 for all the values of $I$ we considered ($N \leq 200$) for all three potentials! 
It must be stated that in order to obtain such accuracies the coefficients 
in the fitting functions are defined with 6-7 significant digits {\it only}. 
It is worth noting that a surprising feature holds for the linear potential: 
\[
\gamma^{(|x|)}_{+} \approx -\gamma^{(|x|)}_{-}\ ,
\] 
see Fig.4 in \cite{BS-1:2021}, with sufficiently high accuracy. 

As a result a conjecture emerges: the property 
\begin{equation}
	\label{I2}
	\gamma\ =\ \gamma(I^2)\ ,
\end{equation}
should hold for any non-singular power-like potential $|x|^m, m>0$. 
{Preliminary calculations for $m=1/2, 3/2, 10$ support this conjecture}.

{\bf 2.}\ Following conjecture (\ref{I2}) let us construct the $1/I$-expansion,
\begin{equation}
	\label{expansion}
	\gamma\ =\ \frac{1}{I}\ {\tilde \gamma} \left(\frac{1}{I^{2}}\right)\ =\ 
	\frac{1}{I}\ \sum_{k=0}^\infty\,\frac{a_k(m)}{I^{2k}}\ , 
\end{equation}
hence, assuming the absence of odd degree terms in the sum. The validity of (\ref{expansion}) 
was checked numerically for the potentials $|x|^{1,4,6}$ by interpolating the $\gamma$ corresponding 
to the energies of the highly excited states with large quantum numbers ranging from 50 to 201 
by using the partial sums of (\ref{expansion}), see below. 

Based on the formalism developed in \cite{Bender:1977},  one can explicitly find the coefficient in front of the leading term in the expansion (\ref{expansion})
\begin{equation}
	\label{a0}
	a_0(m)\ =\ \frac{(m-1)\,\cot \pi/m}{6(m+2)\pi}\ ,\ m > 1\ ,
\end{equation}
where for the case of the harmonic oscillator $a_0(2)=0$. By taking the partial sum of the first four terms in (\ref{expansion}), for the potentials $|x|^{1,4,6}$, we were able to find the following four coefficients in the expansion (\ref{expansion}) by fitting $\gamma$ in the domain $N \in [50, 200]$,
\begin{align*}
	a_0^{(+)}(1) &= -0.039\,403, & a_1^{(+)}(1) &= 0.006\,173, & a_2^{(+)}(1) &= -0.007\,738, & a_3^{(+)}(1) &= 0.023\,819\ , \\
	a_0^{(-)}(1) &= 0.028\,145, & a_1^{(-)}(1) &= -0.006\,627, & a_2^{(-)}(1) &= 0.008\,088, & a_3^{(-)}(1) &= -0.024\,406\ , \\
	a_0(4) &= \frac{1}{12 \pi}, & a_1^{(\pm)}(4) &= -0.002\,763, & a_2^{(\pm)}(4) &= -0.001\,299, & a_3^{(\pm)}(4) &= 0.003\,140\ , \\
	a_0(6) &= \frac{5}{16 \sqrt{3} \pi}, & a_1^{(\pm)}(6) &= -0.003\,298, & a_2^{(\pm)}(6) &= -0.010\,771, & a_3^{(\pm)}(6) &= -0.033\,280\ .
\end{align*}
Each coefficient was found with 6 decimal digits {Interestingly, those first four coefficients do not provide indication to convergence of (8).} Astonishingly, a fit with four 
terms in the partial sum of (8) provides an absolute accuracy $10^{-20}$! In order to reach 
such an accuracy over the whole domain $N \in [0, 200]$ with a fitting function of the type 
of (\ref{g13}), it must be modified by increasing the degrees of the polynomials accordingly,
\[
\gamma\ =\ \frac{{\hat P}_2({I}^2)}{\sqrt{{\hat Q}_5({I}^2)}}\ .
\]
It is worth emphasizing that for $x^{4,6}$-potentials the first four terms of the $1/I$-expansion 
of the form (\ref{expansion}) are the same for positive and negative parity states within six decimal digits. 
This may imply the appearance of exponentially-small terms at large $I$ limit (thus, (\ref{expansion}) should be modified into a trans-series by adding the exponentially-small terms) 
in addition to the Laurent expansion in $1/I$ (\ref{expansion}). This is in the contrast to the situation with the linear potential $|x|$, where the (\ref{expansion}) is different for the positive and negative parity states. However, at the same time another property occurs: to a certain accuracy
\[
a^{(+)}_k(1)\approx -a^{(-)}_k(1)\ .
\]

In a separate study of the quartic oscillator $x^4$ we explored the convergence of the coefficients $a_k(4)$ in (\ref{expansion}) depending on the number of terms in partial sums,
\begin{equation}
	\label{expansion-partial}
	\gamma\ =\ \frac{1}{I}\ {\tilde \gamma} \left(\frac{1}{I^{2}}\right)\ =\ 
	\frac{1}{I}\ \sum_{k=0}^{k_{max}}\,\frac{a_k(m)}{I^{2k}}\ , 
\end{equation}
used for analysis, see Table \ref{table}. It can be seen that with increasing $k_{max}$ the convergence of $a_k(4)$ for $k=0,1,...5$ is sufficiently fast and the difference $a^{(+)}_k(4) - a^{(-)}_k(4)$ tends to zero.

{\bf 3.}\ {\it Unexpected features}.

\begin{figure}[htbp]
	\centering
	\centering
	\includegraphics[width=\textwidth]{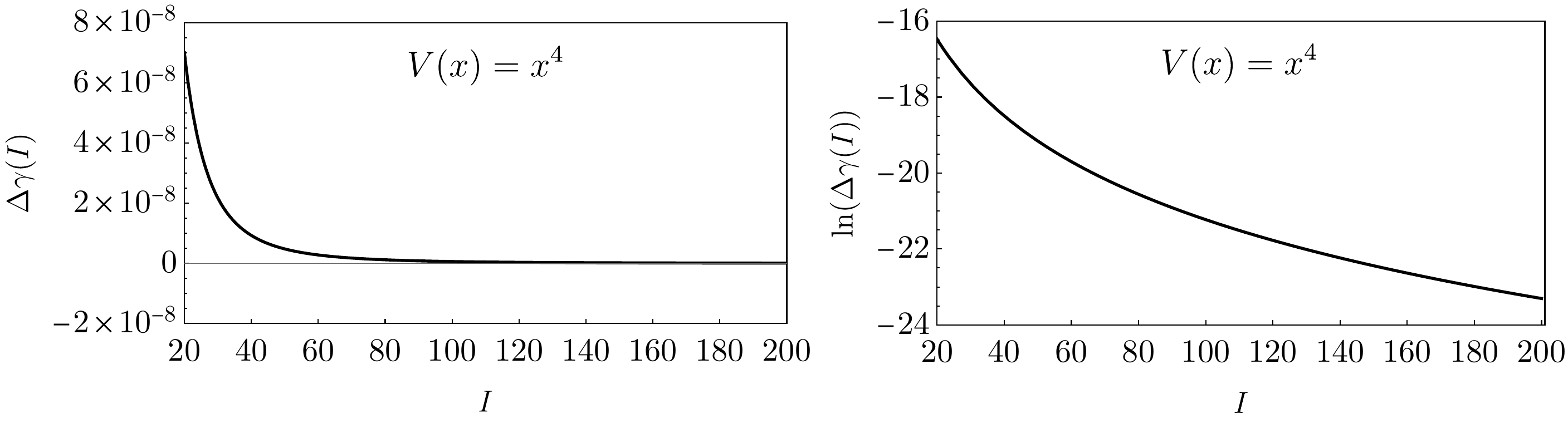}
	\vskip0.25cm
	\caption{Plot of $\Delta \gamma(I)$ for quartic potential $x^4$ as a function of $I = N + 1/2$. 
		The corresponding semi-log plot is shown in the right panel. 
		\label{fig:gamma4}}
\end{figure}

\begin{figure}[htbp]
	\centering
	\includegraphics[width=\textwidth]{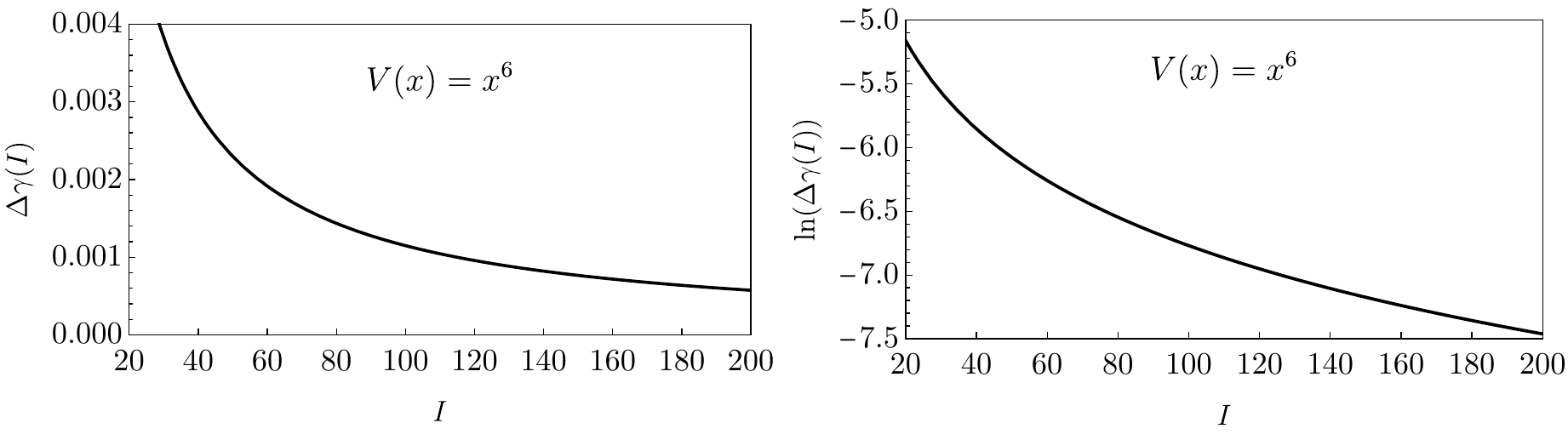}
	\caption{Plot of $\Delta \gamma(I)$ for sextic potential $x^6$ as a function of $I = N + 1/2$. 
		The corresponding semi-log plots is shown in the right panel. 
		\label{fig:gamma6}}
\end{figure}

The plots of the difference $\De \gamma(I)\,=\,\gamma_-(I)-\gamma_+(I)$ for the quartic and sextic potentials, based on the use of (\ref{gamma-quartic-pot}) and (\ref{gamma-sextic-pot}), respectively, is displayed in Fig.\ref{fig:gamma4}, \ref{fig:gamma6}. 
Seemingly, both plots suggest the presence of exponentially small terms at large $I$.
However, even with the accuracy of 200 significant digits in eigenvalues we were unable to find them reliably even though they exist. 

{
}

As for the linear potential $|x|$ let us consider the sum {
	$\Sigma \gamma^{(1)}(I)\,=\,\gamma^{(1)}_-(I)+\gamma^{(1)}_+(I)$, see Fig.\ref{fig:sumgamma}. 
	While both $\gamma^{(1)}_-(I)$ and $\gamma^{(1)}_+(I)$ are described by cumbersome expressions (\ref{gamma-lin-pot}), their sum is small, which is well approximated by a simple formula
	\begin{equation}
		\label{delta1}
		\Sigma \gamma^{(1)}(I)\approx -\frac{0.011\,267 I^2+0.000\,395}{I^3}
	\end{equation}
	\begin{figure}[htbp]
		\includegraphics[scale=0.6]{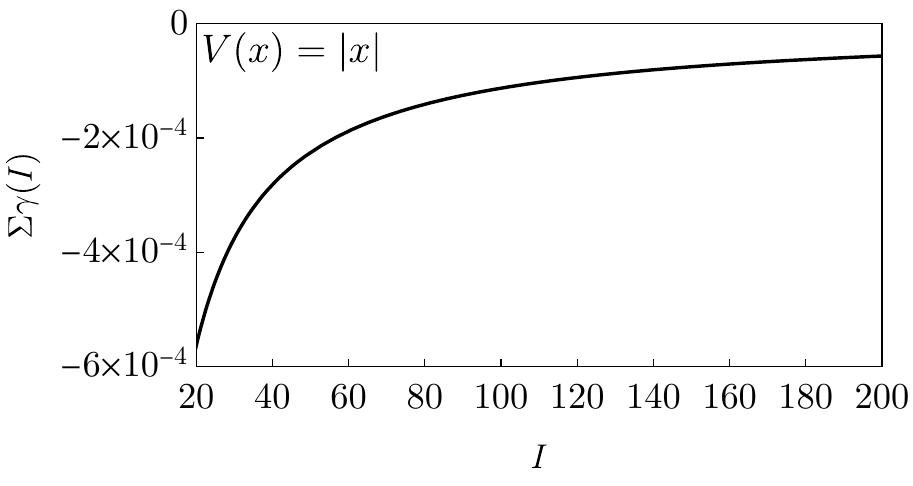}
		\caption{Plot of $\Sigma \gamma(I)^{(1)}$ for the linear potential as a function of 
			$I = N + 1/2$.}  
		\label{fig:sumgamma}
	\end{figure} 
	with accuracy $6\times10^{-6}$, providing an accuracy of at least three  significant digits over the domain $I \in [20, 200]$.
}
\begin{table}
	\centering
	\caption{Quartic Oscillator $x^4$. Absolute errors of the fit of $\gamma$ in the domain $N \in 
		[50,200]$ and coefficients $a_k$ in (\ref{expansion}) for different values of 
		$k_{\text{max}}$ in (\ref{expansion-partial}), shown separately for even (+) and odd (--) states.}
	\resizebox{\textwidth}{!}{
		{\setlength{\tabcolsep}{0.2cm}
			\begin{tabular}{|ccccccccc|}
				\hline\hline
				Parity & $k_{\text{max}}$ & Abs. Error & $a_0$ & $a_1$ & $a_2$ & $a_3$ & $a_4$ & $a_5$  \\
				\hline
				+ & 0 & $1.2 \times 10^{-8}$ & 0.026\,525\,333\,2 & & & & &  \\
				$-$ & 0 & $1.1 \times 10^{-8}$ & 0.026\,525\,349\,1 & & & & &  \\
				\hline
				+ & 1 & $6.1 \times 10^{-13}$ & 0.026\,525\,823\,9 & $-0.002\,763\,474\,1$ & & & & \\
				$-$ & 1 & $5.5 \times 10^{-13}$ & 0.026\,525\,823\,9 & $-0.002\,763\,456\,0$ & & & &  \\
				\hline
				+ & 2 & $1.3 \times 10^{-16}$ & 0.026\,525\,823\,8 & $-0.002\,762\,955\,0$ & $-0.001\,297\,240\,8$ & & &  \\
				$-$ & 2 & $1.2 \times 10^{-16}$ & 0.026\,525\,823\,8 & $-0.002\,762\,955\,0$ & $-0.001\,297\,309\,3$ & & &  \\
				\hline
				+ & 3 & $2.5 \times 10^{-20}$ & 0.026\,525\,823\,8 & $-0.002\,762\,954\,7$ & $-0.001\,299\,179\,1$ & 0.003\,146\,369\,5 & &  \\
				$-$ & 3 & $2.1 \times 10^{-20}$ & 0.026\,525\,823\,8 & $-0.002\,762\,954\,7$ & $-0.001\,299\,179\,0$ & 0.003\,146\,147\,6 & &  \\
				\hline
				+ & 4 & $1.17 \times 10^{-23}$ & 0.026\,525\,823\,8 & $-0.002\,762\,954\,7$ & $-0.001\,299\,177\,4$ & 0.003\,140\,106\,6 & 0.007\,554\,233\,0 &  \\
				$-$ & 4 & $9.2 \times 10^{-24}$ & 0.026\,525\,823\,8 & $-0.002\,762\,954\,7$ & $-0.001\,299\,177\,4$ & 0.003\,140\,105\,5 & 0.007\,555\,665\,2 &  \\
				\hline
				+ & 5 & $7.5 \times 10^{-27}$ & 0.026\,525\,823\,8 & $-0.002\,762\,954\,7$ & $-0.001\,299\,177\,4$ & 0.003\,140\,091\,3 & 0.007\,594\,637\,1 & $-0.038\,877\,577\,7$  \\
				$-$ & 5 & $5.7 \times 10^{-27}$ & 0.026\,525\,823\,8 & $-0.002\,762\,954\,7$ & $-0.001\,299\,177\,4$ & 0.003\,140\,091\,3 & 0.007\,594\,627\,6 & $-0.038\,867\,350\,0$  \\
				\hline\hline
		\end{tabular}}
	}
	\label{table}
\end{table}

{\bf 4.}\ {\it Conclusions.}

For potentials $|x|^{1,4,6}$ we checked numerically that the quantum WKB correction $\gamma$ 
depends on $(N+1/2)^2$, where $N$ is the quantum number. It is conjectured that this property 
should hold for any non-singular power-like potential $|x|^m, m >0$. Whether this property will 
hold for the general polynomial potential, it remains an open problem. For quartic $x^{4}$ and sextic $x^{6}$ potentials it is shown that at large $(N+1/2)$ the $1/|N+1/2|$-expansions for $\gamma$ coincide for the positive and negative parity states while $\gamma^{(+)} \neq \gamma^{(-)}$. It seems this property holds for any power-like potential $|x|^m, m >2$. We failed to find the analytic behavior of the difference $(\gamma^{(+)} - \gamma^{(-)})$ versus the quantum number $N$.

\section*{Acknowledgments}
\vspace{-3mm}

This work is partially supported by DGAPA grant IN104125 (Mexico).


\end{document}